\documentclass[apjl]{emulateapj}
\usepackage{apjfonts}
\begin{document}
\title{The Super-Alfv\'{e}nic Model of Molecular Clouds: Predictions for Mass-to-Flux and
Turbulent-to-Magnetic Energy Ratios}

\author{Tuomas Lunttila\altaffilmark{1}, Paolo Padoan\altaffilmark{2}, Mika Juvela\altaffilmark{1}, and \AA ke Nordlund\altaffilmark{3}}
\altaffiltext{1}{Helsinki University Observatory, P.O. Box 14, T\"{a}htitorninm\"{a}ki, FI-00014,
University of Helsinki, Finland; tuomas.lunttila@helsinki.fi.}
\altaffiltext{2}{Department of Physics, University of California, San Diego, La Jolla, CA
92093-0424; ppadoan@ucsd.edu.}
\altaffiltext{3}{Astronomical Observatory/Niels Bohr Institute, Juliane Maries Vej 30, DK-2100,
Copenhagen, Denmark.}

\begin{abstract}

Recent measurements of the Zeeman effect in dark-cloud cores provide important tests for theories of cloud dynamics and prestellar core formation. In this Letter we report results of {\it simulated} Zeeman measurements, based on radiative transfer calculations through a snapshot of a simulation of supersonic and super-Alfv\'{e}nic turbulence. We have previously shown that the same simulation yields a relative mass-to-flux ratio (core versus envelope) in agreement with the observations (and in contradiction with the ambipolar-drift model of core formation). Here we show that the mass-to-flux and turbulent-to-magnetic-energy ratios in the simulated cores agree with observed values as well. The mean magnetic field strength in the simulation is very low, $\bar{B}=0.34$~$\mu$G, presumably lower than the mean field in molecular clouds. Nonetheless, high magnetic field values are found in dense cores, in agreement with the observations (the rms field, amplified by the turbulence, is $B_{\mathrm{rms}}=3.05$~$\mu$G). We conclude that a strong large-scale {\it mean} magnetic field is not required by Zeeman effect measurements to date, although it is not ruled out by this work.

\end{abstract}
\section{Introduction}

Supersonic turbulence can naturally lead to the complex structure and the large density contrast found in molecular clouds, a process often referred to as turbulent fragmentation. Magnetic fields can affect the outcome of the turbulent fragmentation by reducing the density contrast of shocks, by providing support against gravitational collapse, and by enhancing angular momentum transfer. To test and calibrate star-formation models and simulations based on observational data we must determine the magnetic field strength in the observed star-forming regions. This can be achieved by measurements of the Zeeman effect on molecular emission lines, but only few such measurements have been obtained in dark cloud cores to date (32 measurements and 9 detections in \citet{Troland+Crutcher08}).

The detection of the Zeeman effect on a handful of molecular cloud cores is not sufficient to provide a direct estimate of the mean or mean-squared magnetic field in star-forming regions, as that would require measuring the field strength at many positions in the same molecular cloud, including low density regions. Furthermore, as discussed in \S3, our radiative transfer calculations show that the OH Zeeman effect gives a magnetic field corresponding to the average value of $\rho^a B_{\rm los}$ along the line of sight, with $a \approx 1.5$. This is because the OH emission in the 1665 MHz and 1667 MHz lines that are used for measuring the magnetic field strength increases with both gas density and excitation temperature, and the excitation temperatures increase with density until the transitions become thermalized at $n \approx 10^3$ cm$^{-3}$. Therefore, even in the rare locations where the Zeeman effect is detected, it does not provide a direct estimate of the mean field in that line of sight, but only the field strength in the densest gas. However, it may be possible to constrain the mean magnetic field strength using numerical simulations, as long as observable quantities derived from the simulations are found to depend on the mean magnetic field.

\citet{Padoan+Nordlund99MHD} discussed several observational tests of numerical simulations of supersonic MHD turbulence that could be used to constrain the mean magnetic field. They concluded that the mean magnetic field in molecular clouds is significantly weaker than previously assumed in models of molecular clouds and theories of star formation, suggesting that molecular-cloud turbulence is super-Alfv\'{e}nic on the average, meaning on scales of few to several parsecs. This super-Alfv\'{e}nic model of star-forming regions was recently used to generate simulated measurements of the Zeeman effect on 18 cm OH lines \citep{Lunttila+08}. It was shown that a super-Alfv\'{e}nic turbulence simulation with the characteristic size, density, and velocity dispersion of star-forming regions could produce dense cores with the same $|B_{\rm los}|$-$N$ relation as observed cores.
Furthermore, \citet{Lunttila+08} computed the relative mass-to-flux ratio ${\cal R_\mu}$, defined as the mass-to-flux ratio of the core divided by that of the envelope, following the observational procedure proposed by \citet{Crutcher+08}. They found a large scatter in the value of ${\cal R_\mu}$, and an average value of ${\cal R_\mu}<1$, in contrast to the ambipolar-drift model of core formation, where the mean magnetic field is stronger and only ${\cal R_\mu}>1$ is allowed. The observational results of \citet{Crutcher+08} confirmed ${\cal R_\mu}<1$ in observed cores, as predicted by \citet{Lunttila+08} for the super-Alfv\'{e}nic model.

In this Letter we present further evidence that the same super-Alfv\'{e}nic simulation compares well with the observational data. We use simulated OH Zeeman measurements to compute the mass-to-flux ratio relative to the critical one, $\lambda$, and the ratio of turbulent to magnetic energies, $\beta_{\rm turb}$, in molecular cores selected from simulated maps. We follow closely the observational procedure of \citet{Troland+Crutcher08}, and find mean values of $\lambda$ and $\beta_{\rm turb}$ in good agreement with their observational results.

\clearpage

\begin{figure*}[t]
\plottwo{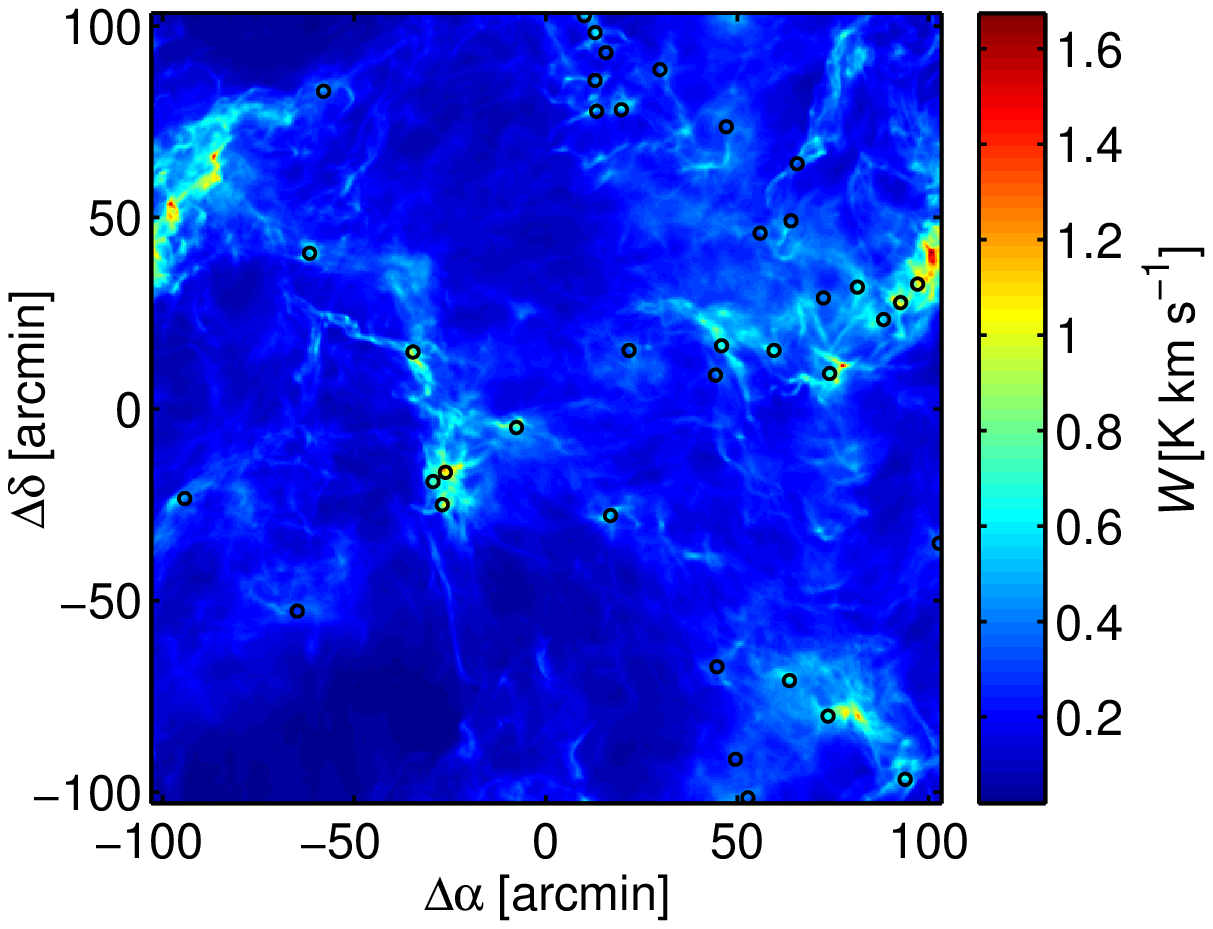}{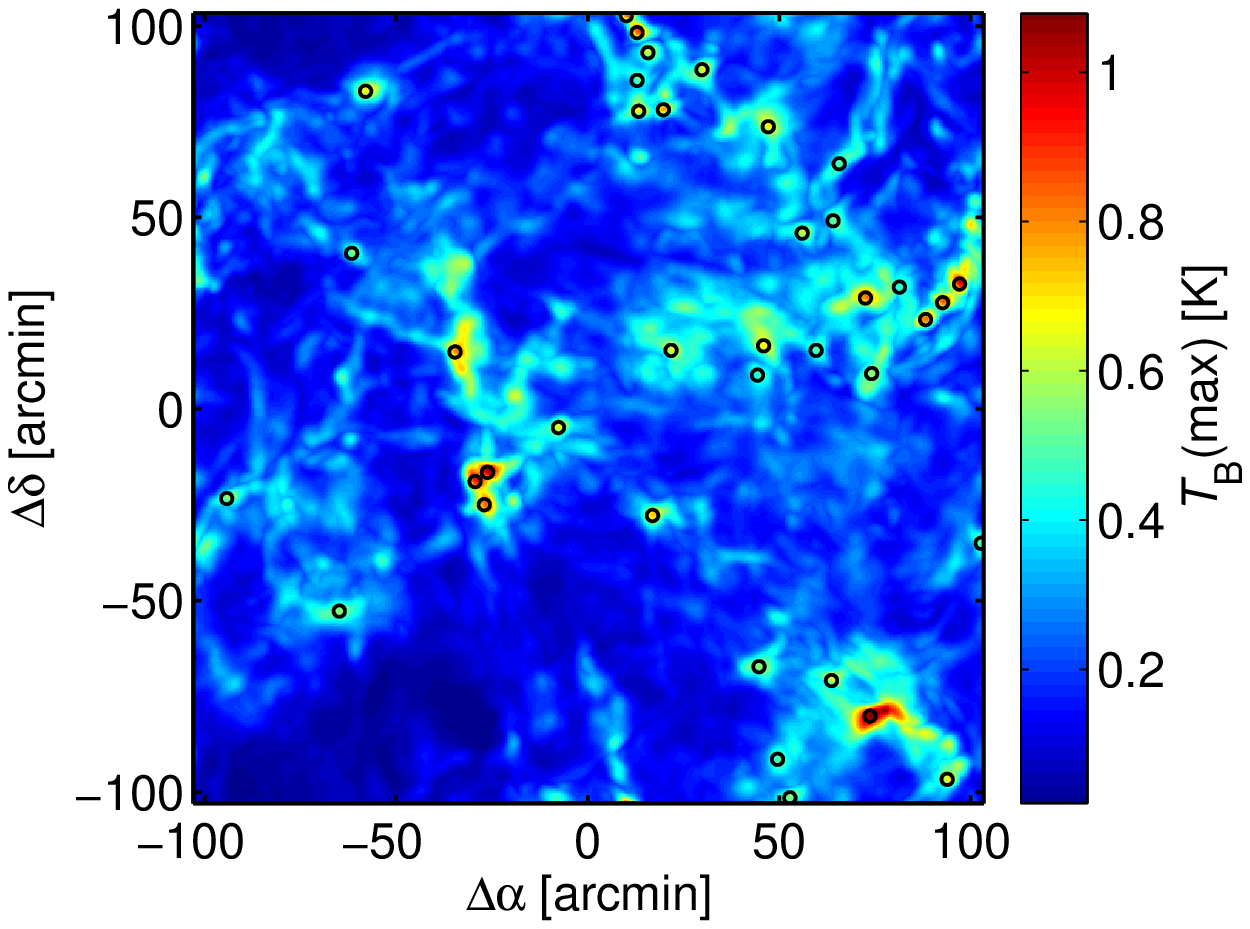}
\caption{\emph{Left:} Simulated 1665 MHz OH integrated intensity map in the y direction, convolved with a $3\arcmin$ beam, assuming the model cloud is at a distance of 150~pc. Circles show the size of the beam (fwhm) and the positions of the cores. \emph{Right:} Core positions shown on the corresponding OH peak brightness temperature map.}
\end{figure*}

\section{Numerical Simulation of Super-Alfv\'{e}nic turbulence}

This work is based on the same simulation of supersonic and super-Alfv\'{e}nic turbulence (and the same snapshot) used in \citet{Lunttila+08}. The simulation was run on a mesh of $1000^3$ zones with the Stagger Code \citep{Padoan+07imf}, with periodic boundary conditions, isothermal equation of state, random forcing in Fourier space at wavenumbers $1\le k\le 2$ ($k=1$ corresponds to the computational box size), uniform initial density and magnetic field, and random initial velocity field with power only at wavenumbers $1 \le k\le 2$.  The rms sonic Mach number is ${\cal M}_{\rm s}=\sigma_{\rm v,3D}/c_{\rm s}= 8.91$.

The initial value of the ratio of gas to magnetic pressure in the simulation is $\beta_{\rm i}= (\rho c_{\rm S}^2) / (B_0^2/8 \pi )= 22.2$, where $c_{\rm S}$ is the sound speed, and $B_0$ is the initial (uniform) magnetic field strength. At the time corresponding to the snapshot used in this work, the rms magnetic field strength has been amplified by the turbulence, giving a value of $\beta=0.2$, defined with the rms magnetic pressure. This corresponds to an rms Alfv\'{e}nic Mach number of ${\cal M}_{\rm a}=(\beta/2)^{1/2}\sigma_{\rm v,3D} / c_{\rm s}=2.8$, so the turbulence is super-Alfv\'{e}nic also with respect to the rms Alfv\'{e}n velocity. With respect to the Alfv\'{e}n velocity corresponding to the mean magnetic field ($\beta_{\rm i}=22.2$), the rms Alfv\'{e}nic Mach number is much larger, ${\cal M}_{\rm a,i}=(\beta_{\rm i}/2)^{1/2}\sigma_{\rm v,3D} / c_{\rm s}=29.7$. Values of parameters scaled to physical units are given in the next section, as the radiative transfer calculations require physical values of size, temperature, and mean density.

\section{Simulated Zeeman Effect}

For the computation of synthetic Zeeman spectra the data cube is scaled to physical units. The size of the grid is fixed to $L=9$ pc, the mean density to $\left<n(\mathrm{H}_2)\right>=67$~cm$^{-3}$ (typical for that scale in the sample of \citet{Falgarone+92}), and the kinetic temperature to $T_{\mathrm{kin}}=10$~K. With this scaling, the mean magnetic field is $(\bar{B}_x,\bar{B}_y,\bar{B}_z)=(0.0, 0.0, 0.34)$~$\mu$G and the rms field $B_{\mathrm{rms}}=3.05$~$\mu$G (much larger than the mean magnetic field, due to the turbulent amplification mentioned above). We assume a constant fractional OH abundance of $\mathrm{[OH]/[H]}=4.0\times 10^{-8}$ from \citet{Crutcher79}.

We simulate Zeeman splitting observations of 1665 and 1667 MHz OH lines that are commonly used for measuring magnetic field strengths in molecular clouds. Full radiative transfer calculations with our line radiative transfer program \citep{Juvela97}, using a datacube resampled to a resolution of $256^3$ cells, showed that the cloud is optically thin and the radiation field is approximately constant throughout the cloud. Thus, the level populations only depend on the local density. Comparison with our full radiative transfer calculations indicate that the errors in the excitation temperatures of the 1665 and 1667 MHz OH transitions are $\la 1$ K. Using the estimated level populations, the coupled radiative transfer equations for the four Stokes parameters are integrated along the line of sight. We use a resolution corresponding to $512^3$ computational cells for computing the synthetic Zeeman spectra.  To increase the number of simulated observations, the calculations are carried out for the x, y, and z directions.

\begin{figure*}[t]
\plottwo{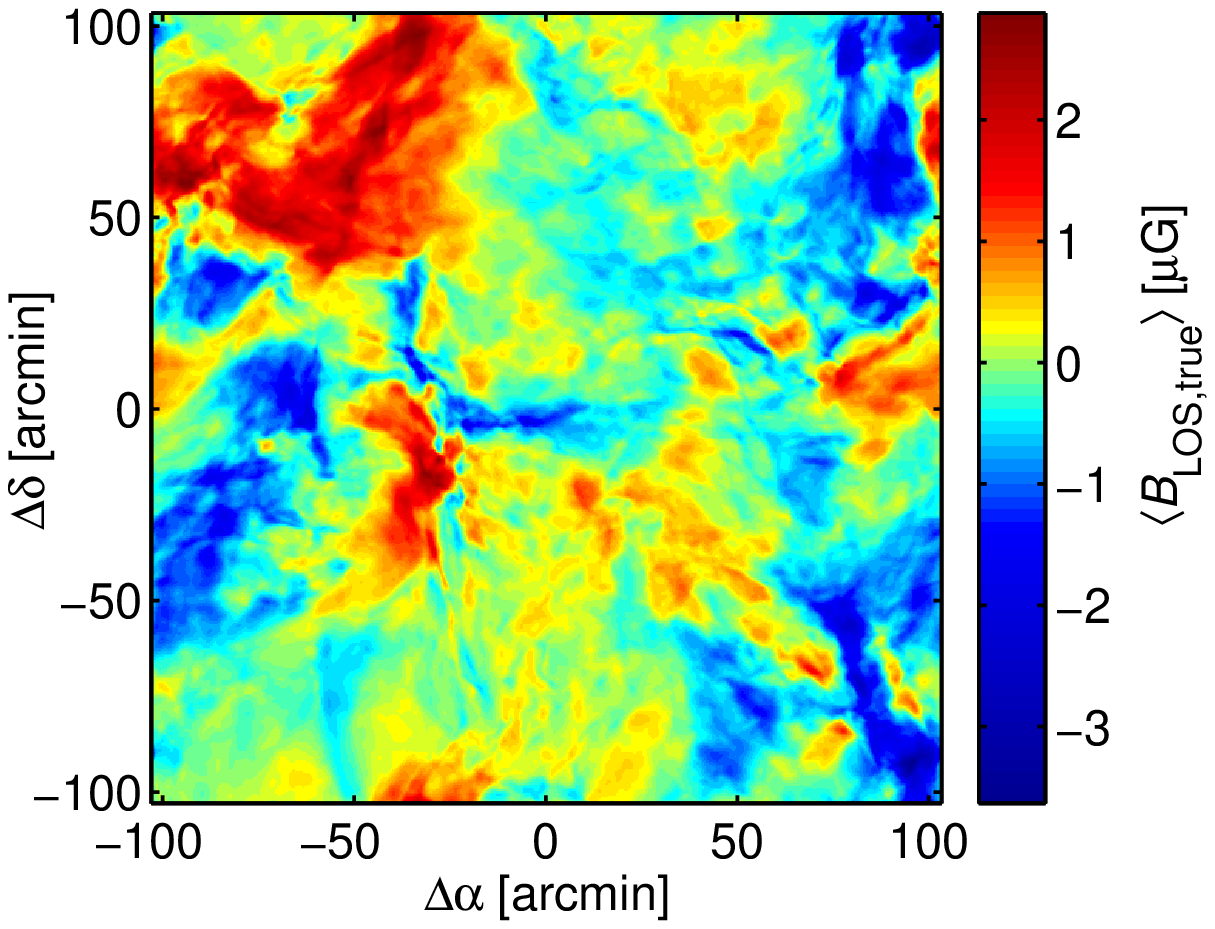}{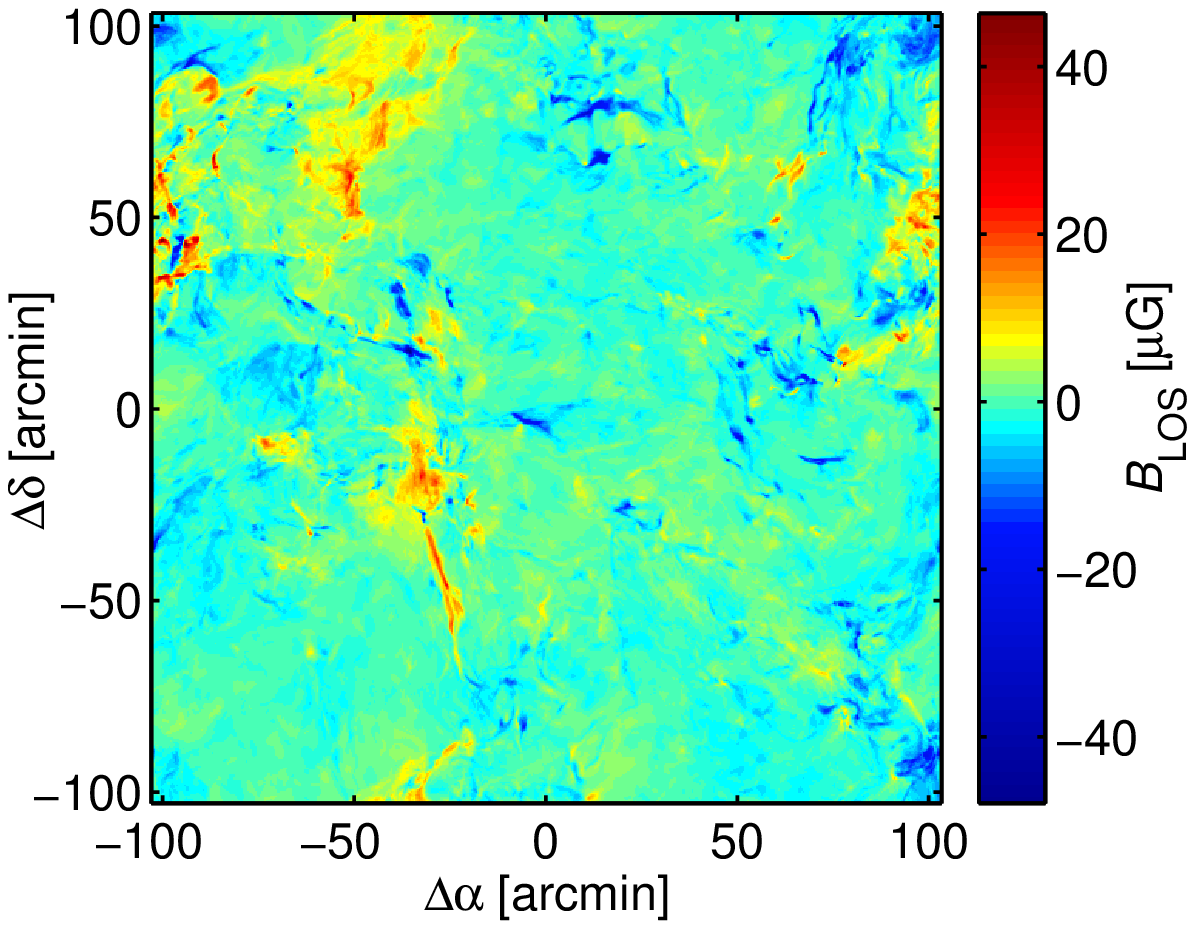}
\caption{\emph{Left:} Mean value of the line-of-sight magnetic field computed directly from the three-dimensional data cube. \emph{Right:} Line-of-sight magnetic field estimated from the simulated Zeeman measurements.}
\end{figure*}

\begin{figure*}[t]
\plottwo{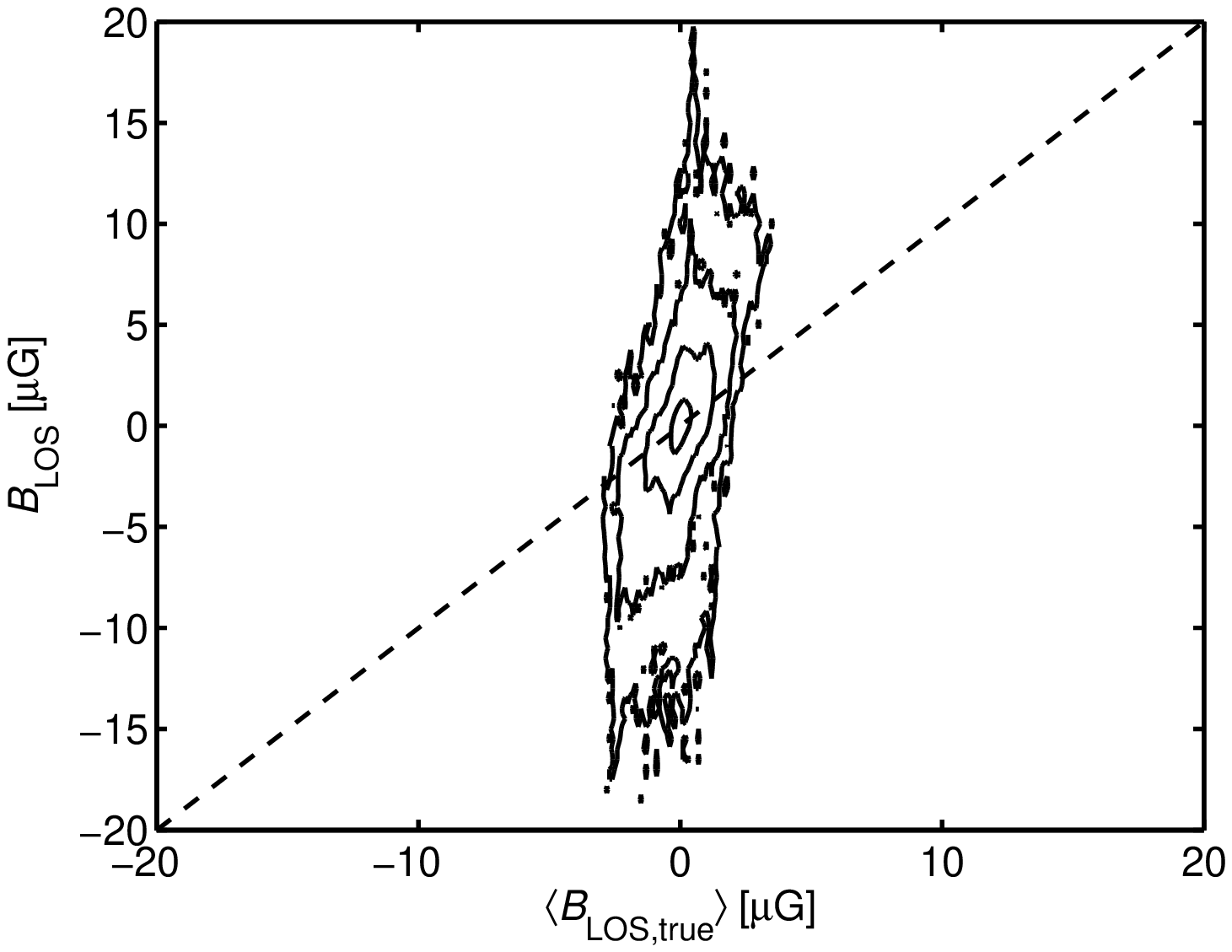}{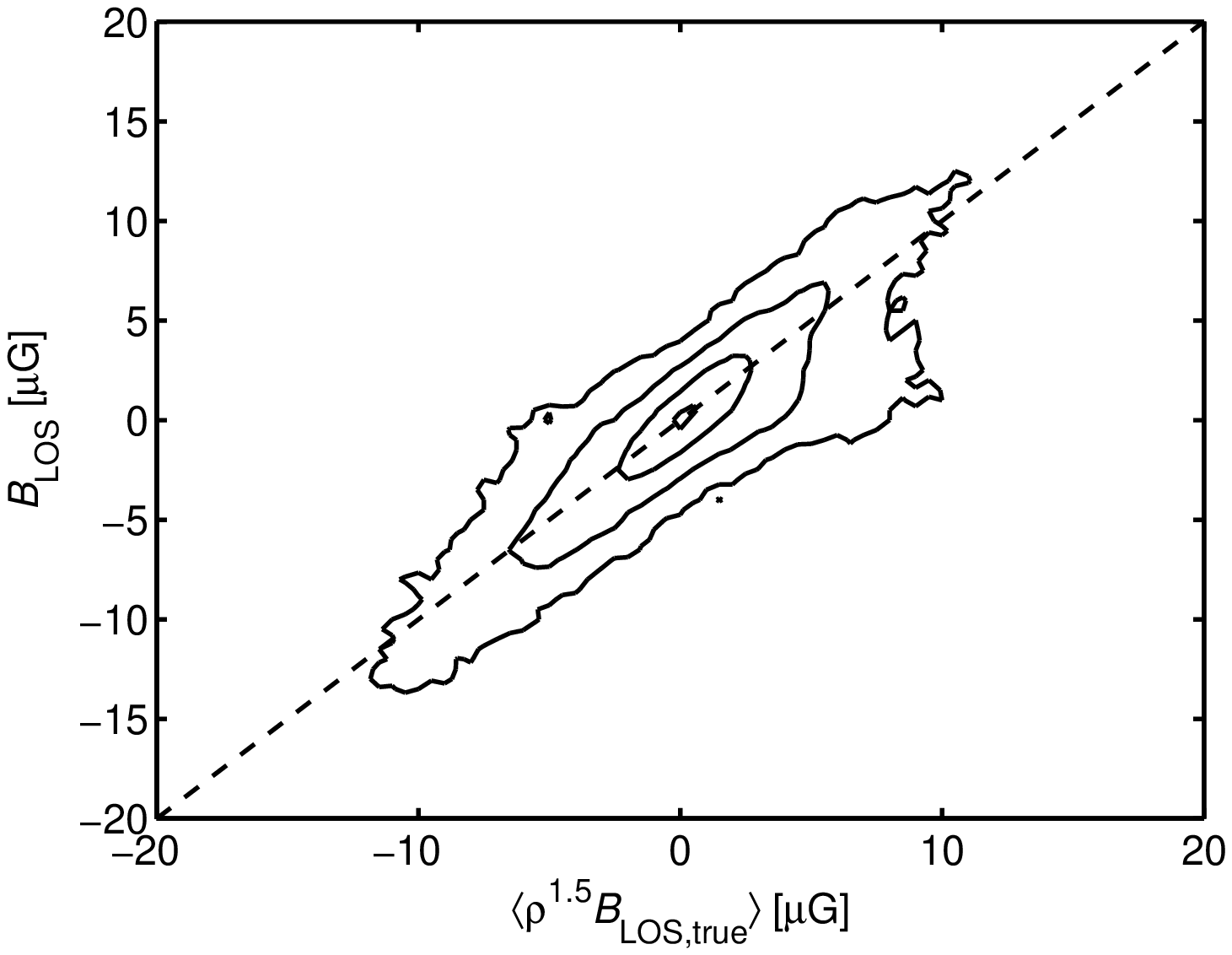}
\caption{\emph{Left:} Isodensity contours of the line-of-sight magnetic field estimated from the Zeeman measurements on the 1665 MHz OH line, versus the mean of the true line-of-sight magnetic field. \emph{Right:} Isodensity contours of the line-of-sight magnetic field estimated from the Zeeman measurements versus the average of the true line-of-sight magnetic field multiplied by $\rho^{1.5}$. The contour levels in both panels are 40, 400, 4000, and 40000 pixels $\mu$G$^{-2}$.}
\end{figure*}

To compare the results with observations, we simulate observations at cloud distances of $D=150$ pc, $D=300$ pc, and $D=1000$ pc, yielding total angular sizes of $\sim 3.4\degr$, $\sim 1.7\degr$, and $\sim 31\arcmin$. We compute position-position-velocity data cubes by simulating observations of the 1665 MHz line. The synthetic observations are made with a $3\arcmin$ (fwhm) beam, corresponding to the angular resolution of the Arecibo telescope, and using a channel separation of $0.05$ km s$^{-1}$. The left panel of Figure~1 shows the simulated 1665 MHz OH integrated intensity map (for the y direction), assuming a cloud distance of $D=150$ pc. The corresponding OH peak brightness temperature map is shown in the right panel of Figure~1.

Figure~2 compares the value of $B_{\mathrm{LOS}}$ estimated from the Zeeman effect (right panel) with the actual mean value of the line-of-sight magnetic field computed directly from the three-dimensional data cube, $\langle B_{\mathrm{LOS,true}}\rangle$ (left panel), for the same map shown in Figure~1. The Zeeman effect tends to select the densest regions along the line of sight, where the magnetic field is stronger (dense cores originate from super-Alfv\'{e}nic compressions that amplify the magnetic field components on the plane perpendicular to the direction of compression). As a result, the Zeeman effect greatly overestimates the mean magnetic field on lines of sight passing through dense cores -- in the y-direction map, $|\langle B_{\mathrm{LOS,true}}\rangle|$ reaches a maximum value of only 3.7~$\mu$G, while $|B_{\mathrm{LOS}}|$ reaches a maximum value of 48~$\mu$G. Because the Zeeman effect is weighted by OH emission, hence by both density and excitation temperature (which increases with density up to $n\approx 1000$~cm$^{-3}$), the estimated magnetic field strength corresponds approximately to the line-of-sight mean value of $\rho^a B_{\mathrm{LOS}}$, with $a \approx 1.5$. The difference between the line-of-sight mean magnetic field and its value estimated through the Zeeman effect is further illustrated in the left panel of Figure~3, showing a scatter plot of $B_{\mathrm{LOS}}$ (from the Zeeman effect) versus $\langle B_{\mathrm{LOS,true}}\rangle$. The right panel of Figure~3 shows that $B_{\mathrm{LOS}}$ is well correlated with $\langle \rho^{1.5}B_{\mathrm{LOS,true}}\rangle$, hence the Zeeman effect is strongly biased towards the magnetic field strength of the densest regions along the line of sight.

We use the computed OH emission position-position-velocity data cubes to select dense cores with the clumpfind algorithm \citep{Williams+94}. Before applying the clumpfind routine, the data cubes are resampled to an angular resolution of $~\sim 1.2\arcmin$ (approximately Nyquist sampled), and uncorrelated Gaussian noise with rms of 0.08 K is added to simulate observational noise. The clumpfind threshold and stepsize parameters are both set to 0.4 K ($5\sigma_{\mathrm{rms}}$). The algorithm provides a list of detected cores and their sizes, which are needed in the subsequent analysis. \citet{Pineda+2009} have recently shown that the clumpfind algorithm is sensitive to the values of the parameters. Running the clumpfind with other parameter values would yield a set of cores with different properties. Nevertheless, our choice of parameters produces a representative ensemble of cores similar to the observed sample of \citet{Troland+Crutcher08}. The core positions are shown as circles on the maps of Figure~1, with the circle size equal to the size of the $3\arcmin$ (fwhm) beam. The cores selected with the clumpfind algorithm do not match very well the column density structure revealed by the integrated intensity map. This is to be expected because the clumpfind algorithm tends to isolate real cores, rather than structures in projection, by selecting distinct features in radial velocity space. As shown by the right panel of Figure~1, all the cores corresponds to local maxima in the OH peak brightness temperature map.

\begin{deluxetable*}{lcrrcrrcrr}
\tablecaption{Physical parameters from the synthetic observations.}
\tablehead{
\colhead{} & & \multicolumn{2}{c}{$D<200$ pc} & & \multicolumn{2}{c}{$200$ pc$<D\le 400$ pc} & & \multicolumn{2}{c}{$D>400$ pc}\\
\colhead{} & & simulations & T\&C08\tablenotemark{1} & & simulations & T\&C08 & & simulations & T\&C08\\
\colhead{} & & $n=105$ & $n=16$ & & $n=40$ & $n=13$ & & $n=4$ & $n=5$}
\startdata
$|B_{\mathrm{LOS}}|$\tablenotemark{a} & & $6.89\pm0.65$ & $6.54\pm1.39$ & & $7.35\pm1.15$ & $11.1\pm2.7$ & & $11.2\pm4.7$ & $17.2\pm10.8$\\
$r$\tablenotemark{b} & & $0.206\pm0.010$ & $0.234\pm0.015$ & & $0.413\pm0.025$ & $0.431\pm0.042$ & & $0.751\pm0.054$ & $1.53\pm0.22$\\
$\Delta V$\tablenotemark{c} & & $0.498\pm0.021$ & $0.622\pm0.063$ & & $0.856\pm0.072$ & $0.890\pm0.110$ & & $1.22\pm0.20$ & $1.67\pm0.10$\\
$N(\mathrm{H}_2)$\tablenotemark{d} & & $3.73\pm0.16$ & $4.32\pm0.56$ & & $3.61\pm0.21$ & $3.78\pm0.35$ & & $3.29\pm0.44$ & $6.58\pm1.00$\\
$n(\mathrm{H}_2)$\tablenotemark{e} & & $3.22\pm0.13$ & $3.09\pm0.40$ & & $1.53\pm0.10$ & $1.48\pm0.13$ & & $0.707\pm0.072$ & $0.765\pm0.125$\\
$M(\mathrm{OH})$\tablenotemark{f} & & $16.5\pm2.2$ & $19.4\pm4.6$ & & $57.1\pm9.1$ & $59.0\pm12.0$ & & $139\pm33$ & $1090\pm210$\\
$M(\mathrm{virial})$\tablenotemark{f} & & $16.9\pm2.2$ & $28.0\pm7.2$ & & $105\pm21$ & $118\pm30$ & & $316\pm108$ & $1070\pm160$\\
$\bar{\lambda}$, $\bar{\lambda}_{\mathrm{TC}}$\tablenotemark{g} & & $4.12^{5.08}_{3.39}$ & $5.04^{8.69}_{3.18}$ & & $3.76^{5.34}_{2.75}$ & $2.63^{4.72}_{1.66}$ & & $2.30^{8.31}_{1.08}$ & $3.30^{15.5}_{1.06}$\\
$\bar{\beta}_{\mathrm{turb}}$, $\bar{\beta}_{\mathrm{turb,TC}}$\tablenotemark{g} & & $1.60^{2.47}_{1.07}$ & $2.78^{8.26}_{1.12}$ & & $2.14^{4.56}_{1.07}$ & $1.07^{3.41}_{0.36}$ & & $0.91^{12.2}_{0.19}$ & $0.97^{21.0}_{0.11}$\\
\enddata
\tablenotetext{1}{Observations from \citet{Troland+Crutcher08}.}
\tablenotetext{a}{Mean line-of-sight magnetic field, $\mu$G.}
\tablenotetext{b}{Mean core radius, pc.}
\tablenotetext{c}{Mean fwhm, km s$^{-1}$.}
\tablenotetext{d}{Mean column density, $10^{21}$ cm$^{-2}$.}
\tablenotetext{e}{Mean volume density, $10^3$ cm$^{-3}$.}
\tablenotetext{f}{Mean mass, $M_\sun$.}
\tablenotetext{g}{$\mathrm{median}_{0.025\,\mathrm{quantile}}^{0.975\,\mathrm{quantile}}$}
\end{deluxetable*}

Each of the detected cores is observed with a $3\arcmin$ beam, targeting the position of peak intensity in the core. The simulated observations consist of Stokes I and V spectra of the 1665 and 1667 MHz lines for each core. To simulate noise in the observations, uncorrelated Gaussian noise is added to both I and V spectra. The line-of-sight magnetic field is determined by least-squares fitting the numerical derivative of the Stokes I spectrum to the Stokes V spectrum, as it is usually done with Zeeman splitting observations \citep[e.g.][]{Crutcher+93, Bourke+2001}. The magnetic field strength is determined separately from both the 1665 MHz and the 1667 MHz lines, and the analysis uses the average of the two measurements weighted by their inverse squared errors in $B$.

Column densities are estimated from the simulated spectra. Assuming that the line is optically thin, the column density of OH [cm$^{-2}$] is obtained as $N(\mathrm{OH})=a(1-T_{\mathrm{bg}}/T_{\mathrm{ex}})^{-1}W$, where $W$ is the integrated line area [K km s$^{-1}$], $T_{\mathrm{bg}}$ is the background continuum brightness temperature (2.73 K), $T_{\mathrm{ex}}$ is the excitation temperature of the transition, and $a$ is $4.04\times 10^{14}$ for the 1665 MHz line and $2.24\times 10^{14}$ for the 1667 MHz line \citep{Crutcher79}. We use $T_{\mathrm{ex}}=10$ K in the analysis of our synthetic observations. The total H$_2$ column density is obtained as $N(\mathrm{H}_2)=N(\mathrm{OH})/8\times 10^{-8}$. Core masses and volume densities are calculated with the same formulas as in \citet{Troland+Crutcher08}: $n(\mathrm{H}_2)=N(\mathrm{H}_2)/2r$, $M(\mathrm{OH})=\pi r^2 N(\mathrm{H}_2)2.8 m_{\mathrm{H}}$, where $r$ is the core radius and $m_{\mathrm{H}}$ is the mass of an $\mathrm{H}$ atom.

\section{Results}
In the synthetic maps, the clumpfind algorithm finds a total of 105 cores at $D=150$ pc, 40 at $D=300$ pc and four at $D=1000$ pc. Our simulated observations of the cores use a noise level corresponding to a 30-hour integration with a system temperature of 35 K, similar to parameters in recent OH Zeeman surveys~\citep[e.g.][]{Bourke+2001,Troland+Crutcher08}. As in actual Zeeman splitting observations, due to very low amplitude of the Stokes V spectrum, statistically significant detections of the magnetic field are rare. With the commonly used criterion, $|B_{\mathrm{LOS}}|>3\sigma_{B(\mathrm{LOS})}$, there are only 24 detections at $D=150$, four at $D=300$ pc and no detections at 1000 pc. The ratio of the number of $3\sigma$ detections to the number of observed cores, 28/149, is similar to what has been achieved in Zeeman surveys~\citep[e.g.][]{Crutcher+93,Bourke+2001,Troland+Crutcher08}.

In \citet{Troland+Crutcher08}, the authors employ an additional criterion for determining whether the measurement is a detection: 1665 MHz and 1667 MHz lines must yield the same $B_{\mathrm{LOS}}$ within the measurement uncertainties. If the same threshold is used as in \citet{Troland+Crutcher08}, $\Delta B_{\mathrm{LOS}}=|B_{\mathrm{LOS}}(1665)-B_{\mathrm{LOS}}(1667)|<1.9\sigma_{\Delta B}$, three of the 28 $3\sigma$ detections fail this criterion. \citet{Troland+Crutcher08} also use a subjective criterion that the V spectrum must look consistent with detection of the Zeeman effect. This is the case for all 28 $3\sigma$ detections in our simulations. However, in some of these spectra (and in some of the observed spectra as well \citep[e.g.][]{Bourke+2001,Troland+Crutcher08}) there are signs of multiple velocity components with differing line-of-sight magnetic fields.

\citet{Troland+Crutcher08} observed cores at distances ranging from 140 pc to 2000 pc. Because the $3\arcmin$ beam selects structures at very different scales at the extremes of this distance range, we compare the simulation results at each assumed distance only to the observations of cores at similar distances. Thus, we consider separately the \citet{Troland+Crutcher08} observations of cores at $D\le200$ pc, 200$<D\le 400$ pc, and $D>400$ pc. The mean values of physical parameters of the cores as determined from the synthetic observations are given in Table 1. For comparison, we also list the corresponding values calculated from the cores in \citet{Troland+Crutcher08}. The results for key physical parameters determined from our synthetic observations are consistent with the results from \citet{Troland+Crutcher08} for the two closest distances. The most distant cores observed by \citet{Troland+Crutcher08} have radii approximately twice larger than the cores in our simulations; they are also approximately 8 times more massive and their linewidths are approximately $\sqrt{2}$ times larger than the simulated cores.

\begin{figure*}[t]
\plotone{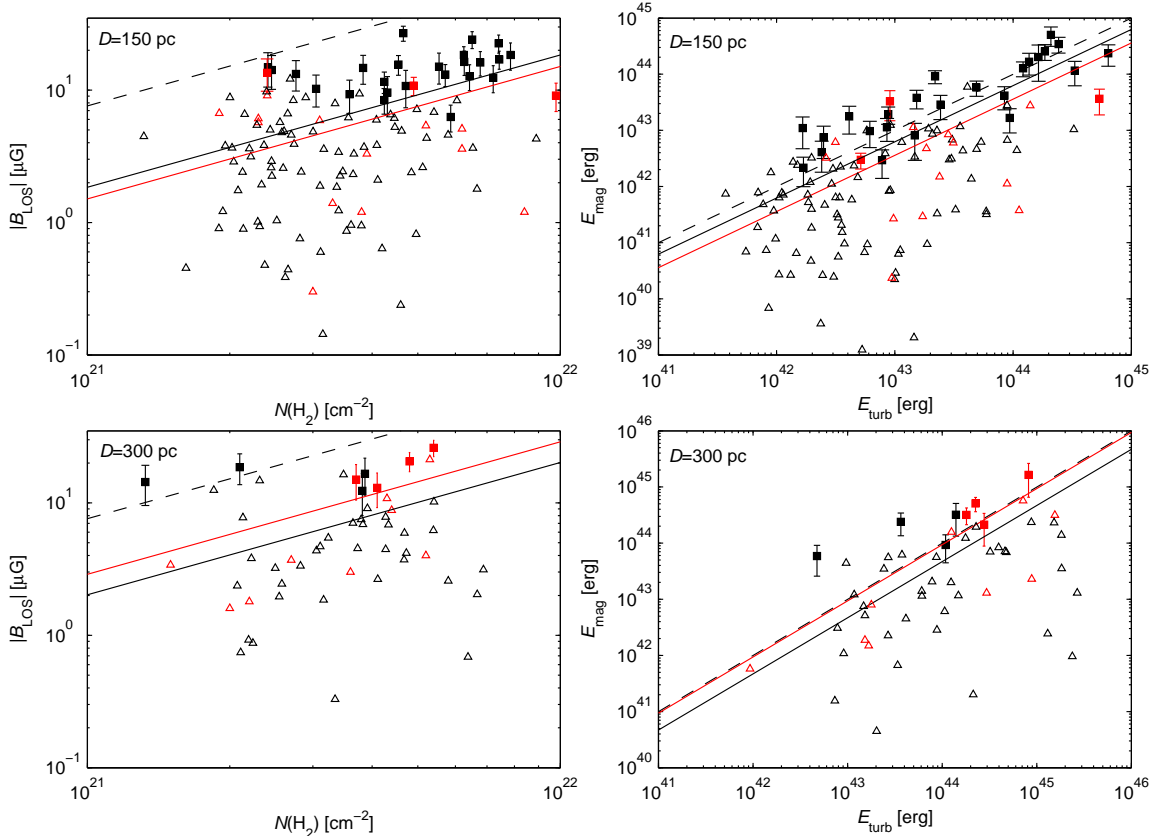}
\caption{\emph{Left:} Inferred line-of-sight magnetic field as a function of H$_2$ column density. Black symbols show the results from our simulation, red symbols are cores from \citet{Troland+Crutcher08}. Squares show $3\sigma$ detections with $\pm 1\sigma$ errorbars, and triangles are non-detections. The black solid line shows the mean mass-to-flux ratio, $\bar{\lambda}$, and the red solid line the mean mass-to-flux ratio calculated from \citet{Troland+Crutcher08} observations, $\bar{\lambda}_{\mathrm{TC}}$. The dashed line is the critical mass-to-flux ratio with no geometrical correction, $\lambda=1$. \emph{Right:} Magnetic energy versus turbulent kinetic energy for the same cores and with the same symbols as in the left panel. The dashed line corresponds to $\beta_{\mathrm{turb}}=E_{\mathrm{turb}}/E_{\mathrm{mag}}=1$. The solid black line shows the mean energy ratio, $\bar{\beta}_{\mathrm{turb}}$, and the red solid line the mean energy ratio calculated from \citet{Troland+Crutcher08} observations, $\bar{\beta}_{\mathrm{turb,TC}}$. The upper panels compare the results of simulated observations at $D=150$ pc with observations of cores at $D\le 200$ pc. The lower panels are for simulated cores at $D=300$ pc and observations in the range 200 pc $<D\le 400$ pc.}
\end{figure*}

Following \citet{Troland+Crutcher08}, we define $\lambda=(M/\Phi)_{\mathrm{obs}}/(M/\Phi)_{\mathrm{crit}}$, the ratio of observed and critical mass-to-flux ratios (above the critical mass-to-flux ratio the magnetic field cannot support the cloud against gravitational collapse). The theoretically determined critical mass-to-flux ratio from \citet{Nakano+Nakamura78} yields $\lambda=7.6\times 10^{-21}N(\mathrm{H}_2)/|B_{\mathrm{LOS}}|$, where $N(\mathrm{H}_2)$ is in cm$^{-2}$ and $B_{\mathrm{LOS}}$ is in $\mu$G. Because only the line-of-sight component of the magnetic field is known, values of $\lambda$ are not meaningful for individual clouds. Instead, the values given in Table 1 are calculated based on the mean values for $N(\mathrm{H}_2)$ and $|B_{\mathrm{LOS}}|$. Because $\lambda$ is a nonlinear function of the observed parameters, its value and uncertainty are estimated using Monte Carlo sampling. The resulting error distribution has a long tail towards high values of $\lambda$. Furthermore, \citet{Troland+Crutcher08} define the ratio of turbulent and magnetic energies, $\beta_{\mathrm{turb}}=E_{\mathrm{turb}}/E_{\mathrm{mag}}$, as $\beta_{\mathrm{turb}}=0.32n(\mathrm{H}_2)\Delta V_{\mathrm{NT}}^2/(3B_{\mathrm{LOS}}^2)$, where $n(\mathrm{H}_2)$ is in cm$^{-3}$, $B_{\mathrm{LOS}}$ is in $\mu$G, and $\Delta V_{\mathrm{NT}}=\sqrt{\Delta V^2-0.027\,\mathrm{km^2\,s^{-2}}}$ is the non-thermal linewidth in km s$^{-1}$. As with the case of $\lambda$, instead of calculating the ratio for individual clouds, we use the mean values of $n(\mathrm{H}_2)$, $\Delta V$, and $B_{\mathrm{LOS}}$, and estimate the error distribution with Monte Carlo sampling. The values of $\bar{\lambda}$, $\bar{\lambda}_{\mathrm{TC}}$, $\bar{\beta}_{\mathrm{turb}}$, and $\bar{\beta}_{\mathrm{turb,TC}}$ in Table 1 are the median values computed from the Monte Carlo sampling around the mean values of the observed parameters.

The mean values of all the observed and derived parameters from the simulation are consistent with those from the sample of \citet{Troland+Crutcher08}. As a result, the values of $\lambda$ from the simulation, $\bar{\lambda}$, are also consistent with the corresponding values from the observations of \citet{Troland+Crutcher08}, $\bar{\lambda}_{\mathrm{TC}}$. The value of $\bar{\lambda}$ is slightly smaller than $\bar{\lambda}_{\mathrm{TC}}$ for cores at distances $D<200$ pc and $D>400$ pc, and slightly larger than $\bar{\lambda}_{\mathrm{TC}}$ for cores at distances in the range 200 pc $< D \le 400$ pc. The same applies to the comparison of the mean energy ratios from the simulation, $\bar{\beta}_{\mathrm{turb}}$, with that from the observations, $\bar{\beta}_{\mathrm{turb,TC}}$. In the case of $\beta_{\mathrm{turb}}$, however, the differences are larger because this ratio depends on the square of the measured linewidth and magnetic field strength. The values of $|B_{\mathrm{LOS}}|$ and $N(\mathrm{H}_2)$ are shown in the left panels of Figure~4 for all cores at distances $D<200$ pc (upper panel) and 200 pc $< D \le 400$ pc (lower panel). The right panels of that figure show the values of $E_{\mathrm{mag}}=4\pi r^3 B_{\mathrm{LOS}}^2/(8\pi)$ and $E_{\mathrm{turb}}=(3/2)M(\mathrm{OH})[\Delta V_{\mathrm{NT}}/(2\sqrt{2\log 2})]^2$ for the same cores. In both plots, the cores selected from the simulation (black symbols) cover approximately the same regions of parameter space as the observed cores (red symbols).

\section{Conclusions}
This work shows that supersonic and super-Alfv\'{e}nic turbulence generates dense cores with physical properties similar to those of observed cores. The mean values of the mass-to-flux ratio and of the turbulent-to-magnetic-energy ratio measured in the simulated cores are consistent with the observed values. As previously shown by \citet{Crutcher+08}, the observed relative mass-to-flux ratio (cores versus envelope) is consistent with the prediction of super-Alfv\'{e}nic turbulence presented in \citet{Lunttila+08} as well. The simulation used in this work has a very low mean magnetic field strength, ${\bar B}=0.34$~$\mu$G, presumably lower than in real molecular clouds. We have chosen this extreme example of super-Alfv\'{e}nic turbulence to show that the observed core magnetic properties do not require a large mean magnetic field on the large scale. The reason why a field strength much larger than the mean may be found in dense cores, is that the field is locally amplified by the turbulence, particularly by the strong compressive motions creating the cores.

We conclude that all Zeeman measurements to date are consistent with the super-Alfv\'{e}nic model of molecular cloud turbulence and core formation proposed by \citet{Padoan+Nordlund99MHD}. As shown by \citet{Crutcher+08}, the observations also appear to contradict a model of core formation based on large-scale magnetic support and quasi-static evolution controlled by ambipolar-drift. While we have proved that a strong large-scale {\it mean} magnetic field is not required by the Zeeman effect measurements, a strong mean field cannot yet be ruled out in the context of supersonic turbulence. In the presence of supersonic turbulence, however, ambipolar drift is not the dominant mechanism of core formation, because cores can be formed dynamically by random compressions along the field.

\acknowledgements{We thank the referee, Alyssa Goodman, for her comments that helped us to improve the paper. This research was supported in part by the NASA ATP grant NNG056601G, NSF grant AST-0507768, a grant from the Danish Natural Science Research Council, the Academy of Finland grants 124620 and 105623, and the V\"{a}is\"{a}l\"{a} foundation. We utilized computing resources provided by the San Diego Supercomputer Center, by the NASA High End Computing Program, and by the Danish Center for Scientific Computing.}

\bibliographystyle{apj}
\bibliography{Zeeman2,MC,padoan}

\begin{thebibliography}{13}
\expandafter\ifx\csname natexlab\endcsname\relax\def\natexlab#1{#1}\fi

\bibitem[{{Bourke} {et~al.}(2001){Bourke}, {Myers}, {Robinson}, \&
  {Hyland}}]{Bourke+2001}
{Bourke}, T.~L., {Myers}, P.~C., {Robinson}, G., \& {Hyland}, A.~R. 2001, ApJ,
  554, 916

\bibitem[{{Crutcher}(1979)}]{Crutcher79}
{Crutcher}, R.~M. 1979, \apj, 234, 881

\bibitem[{{Crutcher} {et~al.}(2009){Crutcher}, {Hakobian}, \&
  {Troland}}]{Crutcher+08}
{Crutcher}, R.~M., {Hakobian}, N., \& {Troland}, T.~H. 2009, \apj, 692, 844

\bibitem[{Crutcher {et~al.}(1993)Crutcher, Troland, Goodman, Heiles, Kaz\`{e}s,
  \& Myers}]{Crutcher+93}
Crutcher, R.~M., Troland, T.~H., Goodman, A.~A., Heiles, C., Kaz\`{e}s, I., \&
  Myers, P.~C. 1993, ApJ, 407, 175

\bibitem[{Falgarone {et~al.}(1992)Falgarone, Puget, \&
  P\'{e}rault}]{Falgarone+92}
Falgarone, E., Puget, J.~L., \& P\'{e}rault, M. 1992, A\&A, 257, 715

\bibitem[{Juvela(1997)}]{Juvela97}
Juvela, M. 1997, A\& A, 322, 943

\bibitem[{{Lunttila} {et~al.}(2008){Lunttila}, {Padoan}, {Juvela}, \&
  {Nordlund}}]{Lunttila+08}
{Lunttila}, T., {Padoan}, P., {Juvela}, M., \& {Nordlund}, {\AA}. 2008, \apjl,
  686, L91

\bibitem[{{Nakano} \& {Nakamura}(1978)}]{Nakano+Nakamura78}
{Nakano}, T., \& {Nakamura}, T. 1978, \pasj, 30, 671

\bibitem[{Padoan \& Nordlund(1999)}]{Padoan+Nordlund99MHD}
Padoan, P., \& Nordlund, {\AA}. 1999, ApJ, 526, 279

\bibitem[{{Padoan} {et~al.}(2007){Padoan}, {Nordlund}, {Kritsuk}, {Norman}, \&
  {Li}}]{Padoan+07imf}
{Padoan}, P., {Nordlund}, {\AA}., {Kritsuk}, A.~G., {Norman}, M.~L., \& {Li},
  P.~S. 2007, \apj, 661, 972

\bibitem[{{Pineda} {et~al.}(2009){Pineda}, {Rosolowsky}, \&
  {Goodman}}]{Pineda+2009}
{Pineda}, J.~E., {Rosolowsky}, E.~W., \& {Goodman}, A.~A. 2009, \apjl, 699,
  L134

\bibitem[{{Troland} \& {Crutcher}(2008)}]{Troland+Crutcher08}
{Troland}, T.~H., \& {Crutcher}, R.~M. 2008, \apj, 680, 457

\bibitem[{Williams {et~al.}(1995)Williams, De~Geus, \& Blitz}]{Williams+94}
Williams, J.~P., De~Geus, E.~J., \& Blitz, L. 1995, ApJ, 428, 693

\end{thebibliography}

\end{document}